\documentclass[%
 reprint,
superscriptaddress,
 amsmath,
 amssymb,
 aps,
]{revtex4-2}

\usepackage{graphicx}
\usepackage{dcolumn}
\usepackage{bm}
\usepackage{hyperref}

\usepackage{physics}
\usepackage{mathrsfs}
\usepackage{amsthm}
\usepackage{mathtools}

\newcommand{\I}{\mathcal{I}}
\newcommand{\R}{\mathcal{R}}
\newcommand{\C}{\mathcal{C}}
\newcommand{\N}{\mathcal{N}}
\newcommand{\Laws}{\mathcal{L}}
\newcommand{\E}{\mathcal{E}}

\newcommand{\T}{\mathcal{T}}
\newcommand{\J}{\mathcal{J}}
\newcommand{\F}{\mathcal{F}}
\newcommand{\A}{\mathcal{A}}
\newcommand{\G}{\mathcal{G}}

\newcommand{\Z}{\mathbb{Z}}
\newcommand{\Real}{\mathbb{R}}
\newcommand{\B}{\mathbb{B}}

\newcommand{\supp}{\mathrm{supp}}

\begin{document}

\preprint{APS/123-QED}

\title{Transitions and Thermodynamics on Species Graphs of\\ Chemical Reaction Networks}

\author{Keisuke Sugie}
\email[E-mail me at: ]{sgekisk@sat.t.u-tokyo.ac.jp}
\affiliation{Department of Mathematical Informatics, Graduate School of Information Science and Technology, The University of Tokyo, Tokyo 113-8654, Japan}

\author{Dimitri Loutchko}
\affiliation{Institute of Industrial Science, The University of Tokyo, 4-6-1, Komaba, Meguro-ku, Tokyo 153-8505 Japan}

\author{Tetsuya J. Kobayashi}%
\affiliation{Department of Mathematical Informatics, Graduate School of Information Science and Technology, The University of Tokyo, Tokyo 113-8654, Japan}
\affiliation{Institute of Industrial Science, The University of Tokyo, 4-6-1, Komaba, Meguro-ku, Tokyo 153-8505 Japan}


\date{\today}

\begin{abstract}
Chemical reaction networks (CRNs) exhibit complex dynamics governed by their underlying network structure.
In this paper, we propose a novel approach to study the dynamics of CRNs by representing them on species graphs (S-graphs). 
By scaling concentrations by conservation laws, we obtain a graph representation of transitions compatible with the S-graph, which allows us to treat the dynamics in CRNs as transitions between chemicals.
We also define thermodynamic-like quantities on the S-graph from the introduced transitions and investigate their properties, including the relationship between specieswise forces, activities, and conventional thermodynamic quantities. 
Remarkably, we demonstrate that this formulation can be developed for a class of irreversible CRNs, while for reversible CRNs, it is related to conventional thermodynamic quantities associated with reactions.
The behavior of these specieswise quantities is numerically validated using an oscillating system (Brusselator).
Our work provides a novel methodology for studying dynamics on S-graphs, paving the way for a deeper understanding of the intricate interplay between the structure and dynamics of chemical reaction networks.
\end{abstract}

\maketitle

\section{Introduction.}
The theoretical framework of chemical reaction network (CRNs for short) is applied in various research fields, such as steady-state flux analysis in metabolic systems \cite{Orth2010} and modeling non-equilibrium processes like oscillatory phenomena 
\cite{doi:10.1073/pnas.0608665104}.
To investigate the properties that hold in a broad range of processes in CRNs, including non-equilibrium states, thermodynamic approaches have recently been used \cite{10.21468/SciPostPhysLectNotes.80, PhysRevLett.127.160601}. Mathematically, CRNs possess a geometric structure represented by hypergraphs with stoichiometric information \cite{Kobayashi2023}, whose structural constraints govern the behaviour of chemical reaction dynamics, for example, the controllability of chemical reactions \cite{himeoka2024stoichiometric}.

For large-scale CRNs, the species graph approach has been utilized.
This is due to the difficulty in intuitively understanding transitions between chemical species over the hypergraph structure, as well as the fact that graph-based characterisations such as small-world and scale-free properties for graphs are relatively well established \cite{Watts1998, 10.1093/oso/9780198805090.001.0001}.
With this background, graph-based studies using species graphs (S-graph) or reaction graphs \cite{Angeli2009} are often preferred over hypergraphs for large networks such as metabolic systems \cite{Wagner2001, Klamt2009}.

However, the dynamics corresponding to the S-graph have not been extensively explored. Despite the potential importance of dynamical oscillations in actual metabolic and glycolytic systems \cite{Himeoka_2016, doi:10.1021/acs.jpcb.1c01325}, the dynamical aspects of species transitions in chemical reaction systems have been neglected.

In this paper, through scaling concentrations by conservation laws, we obtained a graph representation of transitions compatible with the S-graphs of chemical reaction networks. This transition representation enabled us to derive a novel thermodynamics-like formulation for CRNs. Remarkably, we demonstrated that this thermodynamics-like formulation can be applied to a class of irreversible CRNs, while to reversible CRNs, it is reduced to conventional thermodynamic quantities associated with reactions. A key contribution of our research is proposing a methodology to derive a master equation on the S-graph governing the scaled concentrations from deterministic CRN dynamics. Our work provides a novel approach to study of dynamics on S-graphs, paving the way for a deeper understanding of the intricate interplay between the structure and dynamics of chemical reaction networks.

The paper is structured as follows: The following section \ref{Foundation of chemical reaction network.} introduces fundamental terms related to the CRN dynamics and note thermodynamic quantities defined in a reversible CRN. 
In the section \ref{species-transitional representation of stoichiometric dynamics.}, we define a new physical quantity called "transition" and reinterpret the CRN dynamics as dynamics on a species graph. The conservative nature of the dynamics on the graph is realised through concentration scaling by the conservation laws of the network.
In the section \ref{thermodynamics}, we define thermodynamic quantities on the species graph from the transitions we have introduced and investigate their properties. In particular, we discuss the relationship of the specieswise forces and activities with conventional thermodynamic quantities.
The behaviour of the specieswise quantities is checked numerically using an oscillating system (Brusselator) in the section \ref{numerical experiment}.

\begin{figure}[htbp]
    \centering
    \includegraphics[width=.5\textwidth]{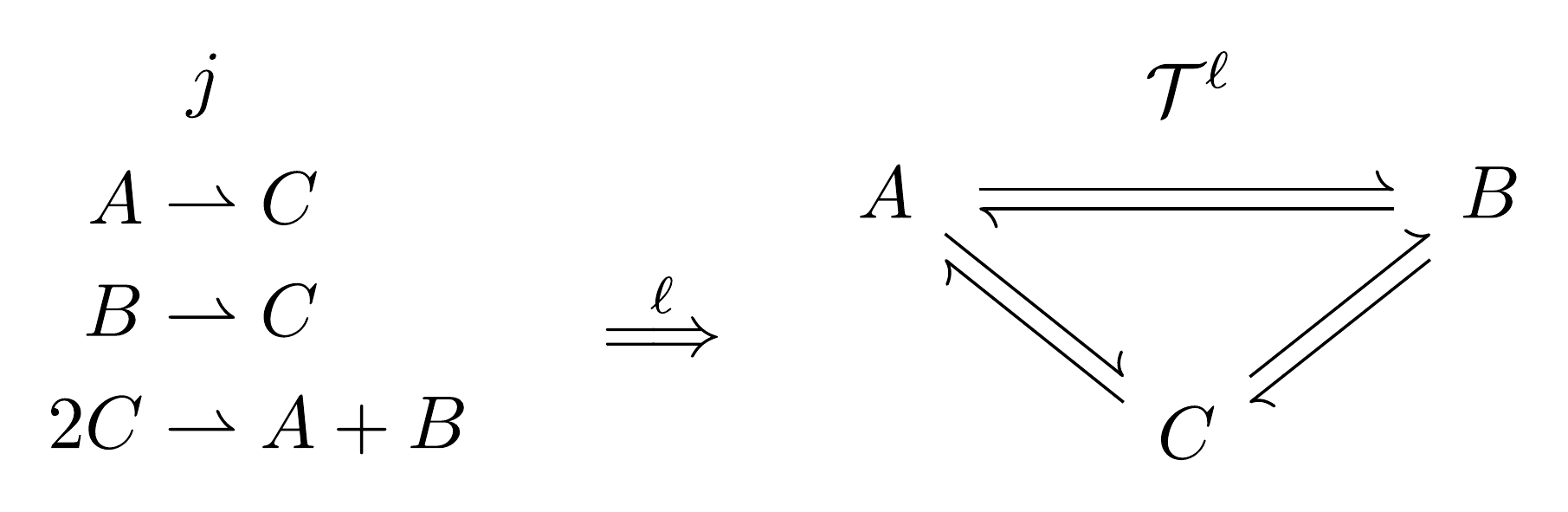}
    \caption{
    Schematic of the method for obtaining the dynamics on the S-graph.
The physical quantity $T^\ell$ (right) compatible with the S-graph is obtained from the CRN dynamics (left), given a conservation law $\ell$ of the CRN.
    }
    \label{fig:force_bound}
\end{figure}

\section{Foundation of chemical reaction network.}
\label{Foundation of chemical reaction network.}
Fundamentals of chemical reaction networks and kinetics are reviewed in this section.
The detailed notations can be found in \cite{Fontanil_Mendoza_2022, Feinberg_2019}.

\subsection{Notation and Preliminaries}
Consider a non empty finite set $\I$ of \emph{species}.
Let $\C \subset \Z_{\geq 0}^\I$ be a finite set of \emph{complexes}.
Suppose a non empty finite set of \emph{reactions} $\R \subseteq \Z_{\geq 0}^\I\times \Z_{\geq 0}^\I$ satisfying the following conditions,
\begin{eqnarray}
    \emptyset &=& \qty{ (y,y) \mid y \in \C } \cap \R,\\
        \C &=& \qty{y, y' \mid (y, y') \in \R}.
\end{eqnarray}
We denote a reaction $(y,y') \in \R$ as $y\to y'$.
For a complex $y \in \C$, its support is defined as $\supp(y) = \qty{i\in \I \mid y_i > 0}$. 
$y$ and $y'$ are respectively called reactant and product complexes of a reaction $y \to y'$.
Each $\supp(y), \supp(y') \subseteq \I$ respectively represents a set of reactants and products.
The triple $\N = (\I,\C,\R)$ of species, complexes and reactions is called a chemical reaction network (CRN) if $\N$ satisfies the above conditions.


We distinguish \emph{proper reactions} $y\to y'$, which have $y \neq 0$ and $y'\neq 0$, \emph{import reactions} for which $y=0$, and \emph{export reactions} for which $y'=0$.
A CRN is \emph{closed} if all reactions $y\to y' \in \R$ are proper.
We focus on a closed CRN $(\I,\C,\R)$, unless explicitly stated otherwise.

Consider $1$-norm $\norm{\cdot}$ on a complex set $\C$ of $\N$;
for a complex $y\in \C$, $\norm{y} \coloneq \sum_{i\in \I} y_i > 0$.
The non-emptiness of a complex's support is equivalent to the positiveness of its norm;  
$\mathrm{supp}(y) \neq \emptyset  \Leftrightarrow  \norm{y} \geq 0$.
In the same way, properness of a reaction is represented in terms of its support or norm,
\begin{eqnarray}
   & & y \to y' \text{ is a proper reaction}\\
    &\Leftrightarrow& \supp(y) \neq \emptyset \text{ and } \supp(y') \neq \emptyset\\
    &\Leftrightarrow& \norm{y}> 0 \text{ and } \norm{y'}> 0.
\end{eqnarray}
In this paper, $\norm{y}$ ($\norm{y'}$) is called a \emph{total reactant (product) coefficient} of a reaction $y\to y' \in \R$.
When a reaction has the same value of its two total coefficients $\norm{y} = \norm{y'}$, the value is simply called a \emph{total coefficient} or \emph{degree} of $y\to y'$.

In our notation, the directed graph $\G_\N$ induced by a CRN $\N = (\I, \C, \R)$ is defined as $\G = (\I, \E)$ with the edge set
\begin{eqnarray}
    \E = \qty{i\gets j \mid \exists y\to y'\in \R_\N, y'_i y_j > 0}
\end{eqnarray}
where the tuple $(i,j)$ is written as $i\gets j$.
$i$ ($j$) is called a \emph{head} (\emph{tail}) of the edge $e = i\gets j$ and denoted as $i = h(e)$ or $j = h(e)$.
The induced graph $(\I, \E)$ is called a \emph{species graph} (or \emph{S-graph} for short), which has a set of species as nodes and edges whose head is a reactant and tail is a product of a certain reaction \cite{Angeli2009, Klamt2009}.
The \emph{incidence matrix} $\B \in \qty{0,\pm 1}^{\I \times \E}$ of the graph $(\I, \E)$ is defined as 
\begin{align}
    \B_{ie} = 
    \begin{cases}
        1 & i \text{ is the tail of } e,\\
        -1 & i \text{ is the head of } e,\\
        0 & \text{otherwise}.
    \end{cases}
\end{align}

\subsection{Conservation laws of CRN}
$\mathcal{S}_\N = \mathrm{span}\qty{ y' - y \mid y\to y' \in\R }$ is called a stoichiometric subspace of $\N$.
Define the orthogonal complementrary subspace of $\mathcal{S}_\N$ for a canonical inner product of $\Real^\I$: $\mathcal{S}^\bot_\N = \qty{ \ell \in \Real^\I \mid \forall z\in \mathcal{S}_\N, \ell^\top z = 0 }$, which is called a set of \emph{conservation laws}.
Each element of $S^\bot_\N$ is called a conservation law of $\N$.
The set of \emph{semi-positive conservation laws}, or non-negative conservation laws, 
of $\N$ is denoted as $\Laws_\N = \mathcal{S}^\bot_\N \cap \Real_{\geq 0}^\I$ \cite{pmid36123755, 10.1371/journal.pone.0100750}.
We define \emph{positive conservation laws}, which is namely an element of the set $\Laws_\N^+ = \mathcal{S}^\bot_\N \cap \Real_{> 0}^\I$. Equivalently, positive laws are semi-positive conservation laws which have no $0$ elements.
A CRN $\N$ is called \emph{conservative} if there exists a nontrivial positive law, i.e. $\Laws_\N^+ \neq \emptyset$.
Denote a CRN $\N$ is \emph{$\ell$-conservative} if $\N$ has a positive conservation law $\ell \in \Laws_\N^+$.
In particular, we write a $1_\I$-conservative CRN simply as a $1$-conservative CRN.

\subsection{Kinetics on chemical reaction network.}
Next we consider the kinetics on a network $\N = (\I,\C,\R)$.
Suppose $\Real_{> 0}^\I$ as a concentration space and assume that the CRN concentration $x_t$ at any time $t$ is in the concentration space, 
which means that the concentration of each species does not go to $0$.
Consider a reaction rate $j: \Real_{> 0}^\I \to \Real_{> 0}^\R$ of $\N$, which is also called kinetics on $\N$.
The deterministic dynamics on the CRN is written as 
\begin{eqnarray}
\label{CRN dynamics}
    \dd{}_t x_t = \sum_{y\to y' \in\R} (y' - y) j_{y\to y'}(x_t).
\end{eqnarray}

The stoichiometric structure restricts the concentration dynamics of the initial condition $x_0$ to a linear subspace $(x_0 + S) \cap \Real_{> 0}^\I$, which is called a stoichiometric compatibility class.
Because of this stoichiometric constraint, we see that the inner product of a conservation law $\ell$ and concentration $x_t$ is conserved throughout the dynamics, i.e. $\ell^\top x_t = \ell^\top x_0$.
The kinetics $j$ is a mass action kinetics if $j_{y\to y'}(x) = k_{y\to y'} x^y = k_{y\to y'} \cdot \prod_{i\in \I} x_i^{y_i}$ with time-independent reaction rate constants $k \in \Real_{> 0}^\R$.
We denote the value $j_{y\to y'}(x)$ as $j_{y\to y'}$ for simplicity.

\subsection{Thermodynamic quantities on reversible CRN.}
We introduce the thermodynamics on a reversible chemical kinetic system.
A CRN $(\I,\C,\R)$ is called \emph{reversible} iff for any reaction $y\to y' \in\R$, the reverse also exists as a reaction: $y'\to y\in\R$.
The reaction set of a reversible CRN $\R$ can be represented as a disjoint sum of two sets $\R^+, \R^-$: $\R = \R^+ \sqcup \R^-$, which are satisfied
\begin{eqnarray}
    y \to y' \in \R^\pm \implies y' \to y \in \R^\mp.
\end{eqnarray}
Reactions of $\R^\pm$ are called forward / backward reactions, respectively.
Some thermodynamic quantities are defined in the reversible CRN system.
$J_{y\to y'} = j_{y\to y'} - j_{y'\to y}$ is called \emph{net current} or \emph{flow} of reaction $y\to y' \in \R^+$.
$F_{y\to y'} = \ln j_{y\to y'} - \ln j_{y'\to y}$ is \emph{thermodynamic force} or \emph{affinity} of $y\to y' \in \R^+$, which indicates the total entropy change in the network and its surroundings.
$A_{y\to y'} = j_{y\to y'} + j_{y'\to y}$ is the quantity called \emph{dynamical activity}, which is the total flow in forward and backward reactions.
The dynamics on the reversible CRN can be rewritten as a linear summation of stoichiometric vectors $y' - y$ for $\to y' \in \R^+$ with coefficients of net currents $J$,
\begin{eqnarray}
    \dd{}_t x &= \sum_{y\to y' \in\R^+} (y' - y) J_{y\to y'}.
\end{eqnarray}

\section{species-transitional representation of stoichiometric dynamics.}
\label{species-transitional representation of stoichiometric dynamics.}
In this section, we provide an insight into CRN dynamics with newly proposed physical quantities.
The quantities represent concentration transitions between chemical species in discrete states.
A CRN uniquely induces a directed graph with chemical species as nodes and transitions between species as edges.

\subsection{Chemical Transitions on Directed Graphs.}

Our first result is the probability-like representation of the CRN dynamics that is followed by the scaled concentration.
Suppose $\ell$ to be a positive conservation law of the CRN.
For the conservation law $\ell$, we call $\ell \circ x$ the $\ell$-scaled concentration of the ordinary concentration vector $x \in X$ of the CRN dynamics and denote as $x^\ell = \ell\circ x \in \Real_{>0}^{\I}$.
Since the sum over $i\in \I$ is conserved, $x^\ell$ can be regarded as a probability disrtibution on $\I$, which follows a deterministic time evolution.
For the concentration following the dynamics of the equation \ref{CRN dynamics}, the time evolution of $\ell$-scaled concentration $x^\ell$ satisfies the following equation:
\begin{eqnarray}
\label{law dynamics}
    \dd{}_t x^\ell_i = \sum_{j \in \I_\ell \setminus\qty{i} } \T_{ij}^\ell - \sum_{j \in \I_\ell \setminus\qty{i}} \T_{ji}^\ell.
\end{eqnarray}
$\T_{ji}^\ell$ is the concentration-dependent \emph{transition} from $i$ to $j$ with respect to the law $\ell$ defined as 
\begin{eqnarray}
    \T_{ij}^\ell = \sum_{\substack{y\to y' \in \R}} \norm{y^\ell}^{-1}  {y'}_i^\ell y_j^\ell j_{y\to y'}(x_t),
\end{eqnarray}
where $\norm{\cdot}$ is $1$-norm and $y^\ell = \ell \circ y$ ($y^\ell = \ell \circ y'$) is the scaled reactant (product) coefficients.
Note that $\norm{y^\ell} = \norm{y'^\ell}$ holds for the two total coefficients of any reaction $y\to y' \in \R$ because $\ell$ is a conservation law.
The equation \ref{law dynamics} can explicitly be rewritten as a continuity-equation-like form  \cite{Kobayashi2023, Grady2010} with the connection matrix $\B$ of the S-graph $(\I, \E)$ as
\begin{align}
\label{continuous eq}
    \dd_t x^\ell = -\B \T^\ell.
\end{align}
In the equation \ref{continuous eq}, $\T^\ell$ is identified with the vector in $\R_{\geq 0}^\E$; $(\T_{ij}^\ell)_{i\gets j \in \E}$, which is justified by the conformity of S-graph edge sets and transitions \ref{edge characterization}.

This equation \ref{law dynamics} is similar to the master equation, but generally this equation alone cannot describe the time evolution of $x^\ell$ because the equation is not closed with respect to $x^\ell$ and not equivalent to the original one \ref{CRN dynamics} in general.
If the law $\ell \not \in \Laws_\N^+$, where $\ell$'s support $\I_\ell = \supp(\ell)$ is not equal to the original set $\I$, the transition matrix $\T^\ell$ depends not only on the $x^\ell$ (i.e. $(x_i)_{i \in \I_\ell}$) but also on the whole concentration $x = (x_i)_{i \in \I}$, and avoiding this kind of argument is the reason why $\ell$ is assumed to be positive (or $\I_\ell = \I$) in the previous paragraph.

It can be verified by a simple calculation that the formula \ref{law dynamics} is correct.
The sum of $\T_{ij}^\ell$ and $\T_{ji}^\ell$ over $j \in \I$ are respectively,
\begin{align}
    \label{first term}
    \sum_{j\in \I} T_{ij}^\ell &= \sum_{y \to y' \in \R} {y_i'}^\ell j_{y\to y'} (x_t), \\
    \label{second term}
    \sum_{j\in \I} T_{ji}^\ell &= \sum_{y \to y' \in \R} y_i^\ell j_{y\to y'} (x_t),
\end{align}
because of $\norm{y^\ell}^{-1} \sum_{j} y^\ell_j = \norm{y'^\ell}^{-1} \sum_{j} {y'}^\ell_j = 1$.
by taking the difference of the above two, \ref{first term} and \ref{second term}, \ref{law dynamics} is proved as follows.
\begin{align}
    \sum_{j\in \I} \T_{ij}^\ell - \sum_{j\in \I} \T_{ij}^\ell
    &= \sum_{y \to y' \in \R} ({y_i'}^\ell - y_i^\ell) j_{y\to y'} (x_t) \\
    &= \ell \circ  \sum_{y \to y' \in \R} ({y_i'} - y_i) j_{y\to y'} (x_t)\\
    &= \ell \circ \dd_t x_i\\
    &= \dd_t x^\ell_i.
\end{align}
The representation \ref{continuous eq} is also proved as
\begin{align}
    \label{cont1}
    \sum_{j\in \I} \T_{ij}^\ell - \sum_{j\in \I} \T_{ij}^\ell
    &= \sum_{e\in \E: h(e) = i} \T_{e}^\ell - \sum_{e\in \E: t(e) = i} \T_{e}^\ell \\
    \label{cont2}
    &= - \qty(\sum_{e\in \E: t(e) = i} \T_{e}^\ell - \sum_{e\in \E: h(e) = i} \T_{e}^\ell) \\
    &= -\B \T^\ell.
\end{align}
The transformation is derived from the compatibility of $\E$ and $\T^\ell$, see \ref{edge characterization}.

\subsection{Description of transition by sampling probability.}
We describe the stochastic aspects of this transition representation of dynamics.
In this subsection, the conservation law is assumed to be a one vector for simplicity $\ell = 1_\I$, where $\norm{y} = \norm{y'}$ holds for any reaction $y \to y'$.
Replacing $y $ by $y^\ell$ and $y'$ by ${y'}^\ell $, a similar argument holds for a general conservation law $\ell$.
The transition from $j$ to $i$ is written as the following form,
\begin{eqnarray}
    \T_{ij} &= \sum_{\substack{y\to y' \in \R}} \mathbb{P}(i \gets j| y\to y') \cdot \norm{y} j_{y\to y'},
\end{eqnarray}
where $\T^{1_\I}$ is denoted as $\T$ and 
\begin{eqnarray}
    \mathbb{P}(i \gets j| y\to y') = \frac{y'_i}{\norm{y'}} \cdot \frac{y_j}{\norm{y}}.
\end{eqnarray}
$j_{y\to y'}(x_t) \dd t$ can be interpreted as the probability that the reaction $y\to y'$ occurs in the interval $[t,t+\dd t]$ under the assumption that reactions occur at most once in the short-time $\dd t$.
$\mathbb{P}(i \gets j| y\to y')$ is proportional to $y'_i \cdot y_j$ where $y_i'$ is $i$'s product coefficient and $y_j$ is $i$'s reactant coefficient of $y\to y'$, and  indicates the chemical species sampling with the conditional probability that the state transitions from $j$ (at $t$) to $i$ (at $t + \dd t$) under the condition that the reaction $y \to y'$ occurs in $[t, t+ \dd t]$.
Thus the joint probability
\begin{eqnarray}
    \begin{split}
            &\mathbb{P}(t+ \dd t,i; t, j; y\to y') \\= &\mathbb{P}(i \gets j| y\to y') \cdot j_{y\to y'}(x_t) \dd t.
    \end{split}
\end{eqnarray}
is defined such that a molecule is in state $j$ at $t$ \& in state $i$ at $t + \dd t$ and $y\to y'$ occurs in $[t,t+\dd t]$. Then the transitions is rewritten as 
\begin{eqnarray}
    \T_{ij} \dd{t} &= \mathbb{E}_{y\to y' \sim \mathbb{P}(t+ \dd t,i; t, j; y\to y')} (\norm{y}),
\end{eqnarray}
which is the expected value of the total product/reactant coefficient for reactions $y\to y'$.
The dynamics of $x$ is
\begin{eqnarray}
    \label{dynamics by probability flow}
    \dd x_i = \sum_{y\to y' \in \R} \norm{y} \sum_{j \in \I} I_{ij},
\end{eqnarray}
where $\dd x_i$ is the small concentration change from $t$ to $t +\dd t$ and $I_{ij}$ indicates the net probablity flow from $j$ to $i$ or the assymetry in the two-time joint probablities, 
\begin{align}
    \begin{split}
                I_{ij} = &\mathbb{P}(t+ \dd t,i; t, j; y\to y') \\- & \mathbb{P}(t+ \dd t,j; t, i; y\to y').
    \end{split}
\end{align}
The equation \ref{dynamics by probability flow} can be interpreted that the change in $x$ is driven by the coefficient-weighted asymmetries in the two-time joint probabilities of each reaction.

\subsection{Reversibility of CRN and species transitions.}
We discuss the reversibility of the CRN structure (or its species graph) and its relation to our transition quantity in this subsection. 
First, we see that the existence of a reaction producing $j$ from $i$ is equivalent to the $ij$ component of the transition matrix $\T^\ell$ being not vanishing.
Derived from this, we study the relationship between the reversibility of CRNs and species graphs and the transitions.

For a positive law $\ell$ and its transition $\T^\ell$, the following two condition is equivalent.
\begin{enumerate}
    \item $\exists y \to y' \in \R$ such that $y_i' y_j > 0$. \label{condition 1}
    \item $\T_{ij}^\ell > 0$. \label{condition 2}
\end{enumerate}
The proof is given as follows.
Assuming the condition \ref{condition 1} and denoting such reaction as $\tilde{y} \to \tilde{y}'$, the positiveness can be found from the expression of $\T^\ell_{ij}$ as 
$\T^\ell_{ij} \geq \norm{\tilde{y}^\ell}^{-1}  \tilde{y}'^\ell_i \tilde{y}_j^\ell j_{\tilde{y}\to \tilde{y}'}(x_t) > 0$, 
using $\tilde{y}_i' \tilde{y}_j > 0$ and $j_{\tilde{y}\to \tilde{y}'}(x) > 0\, (\forall x \in \Real_{>0}^\I)$. Assuming that there are no reactions that satisfies $y_i y_j > 0$, then the transition is always zero: $\T_{ij}^\ell = 0$, and the proof is completed by its contraposition.

Our transition matrix is compatible with the species graph in the following manner.
The edge set of species graph is characterized by the transition $\T^\ell$ as
\begin{align}
    \label{edge characterization}
    \E = \qty{i\gets j \mid \T_{ij}^\ell > 0},
\end{align}
which is checked by the definition of the edge set.
As a corollary, we note that the reversibility of edges in the species graph can be expressed in terms of reversibility with respect to the transition T, that is, the equivalence of the following two conditions:
For a positive law $\ell$ and its transition $\T^\ell$, the following two condition is equivalent.
\begin{enumerate}
    \item $i\gets j \in \E$ is reversible.
    \item $\T_{ij}^\ell > 0 \Leftrightarrow \T_{ji}^\ell > 0$.
\end{enumerate}

We comment on the reversibility of the original CRN and its species graph.
For a reversible conservative CRN, its S-graph is reversible, which is because $y_i' y_j > 0$ for $y\to y' \in \R$ implies $y_j y_i' > 0$ for $y'\to y \in \R$ in reversible CRNs.
However, the converse does not hold in general, i.e. a CRN with the reversible S-graph is not necessarily reversible, which can be checked by the CRN with the following reactions,
\begin{align}
    A &\rightharpoonup C,\\
    B &\rightharpoonup C,\\
    2C &\rightharpoonup A + B.
\end{align}
This CRN has $1$ as a positive conservation law and its transition is positive on the edges
\begin{align}
    \E = \qty{ C\gets A, A\gets C, B\gets A, A\gets B },
\end{align}
which constitutes a reversible S-graph.

\section{Thermodynamical quantities on species graphs.}
\label{thermodynamics}
The distribution on nodes of an reversible graph and its dynamics has been studied in the stochastic  thermodynamics \cite{Seifert_2012}.
Our formulation enables us to treat the CRN dynamics in the style of stochastic thermodynamics.
For simplicity, consider a reversible CRN whose conservation law graphs are undirected, which means that the graphs contain any reverse edges.

We can analogously define some themodynamic quantities for the dynamics on the species graph $\G = (\I, \E)$: specieswise net currents, forces and activities,
\begin{eqnarray}
    \J_{ij}^\ell &=& \T_{ij}^\ell - \T_{ji}^\ell,\\
    \F_{ij}^\ell &=& \ln \T_{ij}^\ell - \ln \T_{ji}^\ell,\\
    \A_{ij}^\ell &=& \T_{ij}^\ell + \T_{ji}^\ell.
\end{eqnarray}
We call these physical quantities \emph{specieswise thermodynamic quantities} of the conservative CRN, to be distinguished from original reaction-related flows $J$, forces $F$ and activities $A$.
The three types of quantities are expressed with the summations of the conventional thermodynamical quantities over the reactions ${y\to y'}$ as 
\begin{eqnarray}
    \J_{ij}^\ell &=&  \sum_{y\to y' \in \R^+} \ell_i \ell_j  \norm{y^\ell}^{-1} (y'_i y_j - y'_j y_i) J_{y\to y'},\\
    \F_{ij}^\ell &=& \ln \frac{\sum_{y\to y' \in \R} \norm{y^\ell}^{-1} y_i' y_j j_{y\to y'}}{\sum_{y\to y' \in \R} \norm{y^\ell}^{-1} y_i' y_j j_{y'\to y}},\\
    \A_{ij}^\ell &=&  \sum_{y\to y' \in \R^+} \ell_i \ell_j  \norm{y^\ell}^{-1} (y'_i y_j + y'_j y_i) A_{y\to y'}.
\end{eqnarray}

In the case of the reversible S-graph, using the net current $\J^\ell$, the dynamics can be written as a form of the continuity equation \cite{Kobayashi2023, Grady2010}, that is,
\begin{align}
    \dd_t x^\ell = -\B_{\E^+} \J^\ell.
\end{align}
$\E^+ \subseteq \E$ is a set of forward edge set which satisfies
$i\gets j \in \E^+ \Leftrightarrow j\gets i \not\in \E^+$ and $\B_{\E^+}$ is the restriction of the S-graph's incidence matrix $\B$ to $\qty{0,\pm 1}^{\I\times \E^+}$.
$\J^\ell$ is identified with the vector $(\J_{ij}^\ell)_{i\gets j \in \E^+}$.

Note that some qualities or inequalities hold between thermodynamic quantities on the graph and ones on the CRN, which give physical meanings of $\J_{ij}^\ell, \F_{ij}^\ell, \A_{ij}^\ell$.



\subsection{Specieswise Forces}
\label{specieswise forces}
Assume that there exists at least one reaction which has $i$ as a product and $j$ as a reactant.
$\F_{ij}^\ell$ is regarded as a quantity representative of thermodynamic forces of reactions which have $j$ as a reactant and $i$ as a product because the inequality below holds for any $y \to y' \in \R_{ij}$,
\begin{eqnarray}
    \label{inequality of F}
    \min_{y\to y'\in \R_{ij}}{F_{y\to y'}} \leq \F_{ij}^\ell \leq \max_{y\to y'\in \R_{ij}}{F_{y\to y'}},
\end{eqnarray}
where $\R^{ij}$ is a set of 
reactions which have $j$ as a reactant and $i$ as a product,
\begin{eqnarray}
    \R_{ij} = \qty{y\to y' \in \R \mid y'_j > 0 \text{ and } y_i > 0}.
\end{eqnarray}
From the first assumption, this reaction subset is non-empty.

The inequality derives from the expression of $\F_{ij}^\ell$ by the reaction set $\R_{ij}$, 
\begin{eqnarray}
    \F_{ij}^\ell = \ln \frac{\sum_{y\to y' \in \R_{ij}} \norm{y^\ell}^{-1} y_i' y_j j_{y\to y'}}{\sum_{y\to y' \in \R_{ij}} \norm{y^\ell}^{-1} y_i' y_j j_{y'\to y}}.
\end{eqnarray}
Each term $\norm{y\to y'}_\ell^{-1} y_i' y_j F_{y\to y'}$ is positive (non-zero), thus the bound of $\F_{ij}^\ell$ \ref{inequality of F} is obtained from the inequality $\min_i a_i / b_i \leq \sum_i a_i / \sum_i b_i \leq \max_i a_i / b_i \, (a_i,b_i > 0)$ and the monotonicity of the logarithmic function $\ln(\cdot)$.
From the derivation, we know that the equality of \ref{inequality of F} holds if and only if the original forces $F_{y\to y'}$ take the same value for each $y\to y' \in\R_{ij}$.

Finally, we mention the asymptotic behaviour of $F_{ij}^\ell$. Consider a situation where a certain reaction kinetics $j_{y\to y'}$ is dominant; $j_{y\to y'} \gg j_{\tilde{y}\to \tilde{y}'}$ for any $\tilde{y}\to \tilde{y}' \in \R_{ij}\setminus \qty{y\to y'}$. In this case, the specieswise force will be approximately the thermodynamic force of the reaction $y\to y'$, because
\begin{align}
\begin{split}
        \F_{ij}^\ell &\simeq \ln \frac{\norm{y^\ell}^{-1} y_i' y_j j_{y\to y'}}{\norm{y^\ell}^{-1} y_i' y_j j_{y'\to y}}\\ &= \ln \frac{j_{y\to y'}}{j_{y'\to y}} = F_{y\to y'}.
\end{split}
\end{align}

\subsection{Specieswise Activities}
\label{specieswise activities}
\begin{figure*}[htbp]
    \centering
    \includegraphics[width=\textwidth]{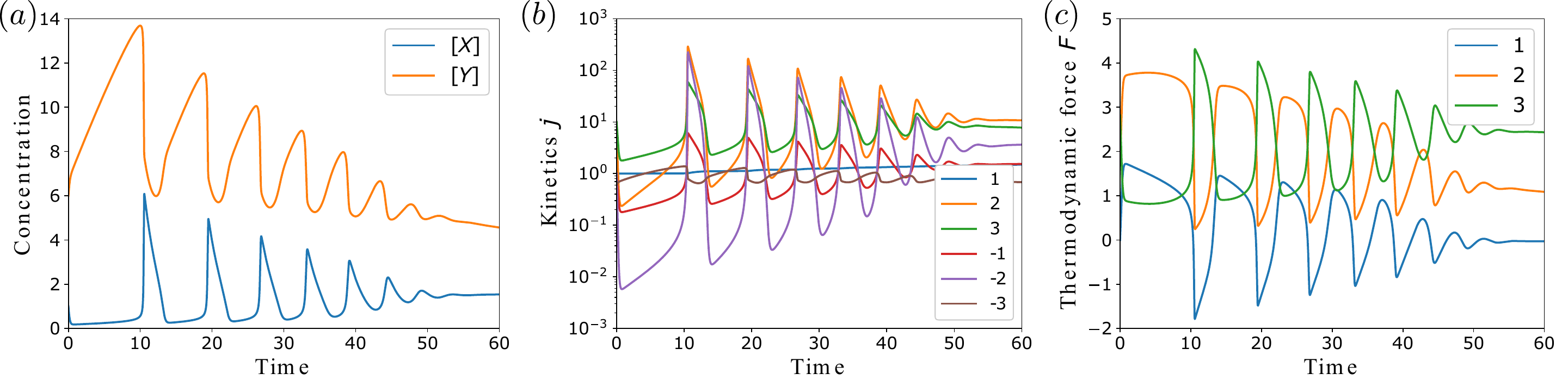}
    \caption{
         The time evolutions of (a) concentrations of species $X$ and $Y$, 
        (b) kinetics $j_m \, (m=\pm 1,\pm 2,\pm 3)$ and 
        (c) thermodynamic forces $F_m \, (m=1,2,3)$.
        The parameters and initial conditions are $k_1^+ = 1\times 10^{-3}, k_1^- = k_2^+ = k_2^- = 1, k_3^+ = 1\times 10^{-2}, k_3^- = 1\times 10^{-4}, [X]_0 = 1, [Y]_0 = 6, [A]_0 = [B]_0 = 1 \times 10^3$ , which is the same as in \cite{PhysRevResearch.3.013175}.
    }
    \label{fig:concentration}
\end{figure*}
Next we see the relationship between the specieswise dynamical activities on the induced graph and the covariance of the chemical dynamics.
The quantity
\begin{eqnarray}
    \label{covariance}
    \tilde{D}_{ij}^\ell = \sum_{y \to y' \in \R^+}((y' - y)^\ell)_i ((y' - y)^\ell)_j A_{y\to y'}
\end{eqnarray}
indicates the (scaled) covariance between $i$ and $j$ of the chemical Fokker-Plank equation \cite{10.1063/1.481811}.

$\A_{ij}^\ell$ gives a lower bound for the covariance $\tilde{D}_{ij}^\ell$:
\begin{eqnarray}
    \tilde{D}_{ij}^\ell \geq - \qty(\max_{y\to y' \in \R^+} \norm{y^\ell} ) \cdot \A_{ij}^\ell.
\end{eqnarray}
The proof is completed by a simple transformation of the equation \ref{covariance} and the process is shown below.
\begin{align}
    \tilde{D}_{ij}^\ell &= \sum_{y \to y' \in \R^+} \qty{( y'_i y_i + y'_j y_j ) - ( y'_i y_j + y'_j y_i )} A_{y\to y'} \\
    &\geq - \sum_{y \to y' \in \R^+}  ( y'_i y_j + y'_j y_i ) A_{y\to y'} \\
    &= - \qty(\max_{y\to y' \in \R^+} \norm{y^\ell} ) \cdot \A_{ij}^\ell.
\end{align}
The equality holds if and only if the total coefficient $\norm{y^\ell} = \norm{y'^\ell}$ takes the same value for any reaction $y\to y' \in \R^+$ and no reaction exists where $i,j$ are both reactants or products.

The asymptotic analysis differs slightly from the force case.
In order for the reaction $y \to y'$ to asymptotically coincide with the dominant case $j_{y\to y'} \gg j_{\tilde{y}\to \tilde{y}'}$, an additional condition is required such that this reaction must be of maximum order throughout the entire reaction system: $\norm{y^\ell} = \max_{\tilde{y} \to \tilde{y}' \in \R_{ij}} \norm{\tilde{y}^\ell}$.

\section{numerical experiment}
\label{numerical experiment}

Through a conservative CRN example, we confirm the behaviour of our defined physical quantities.
The Brusselator is used for numerical experiment, which is a well-known autocatalytic reaction model showing chemical oscillation \cite{lefever1988brusselator}.
We use the damped version of Brusselator system in \cite{PhysRevResearch.3.013175} which has $4$ species $\I = \qty{A, B, X, Y}$ and $6$ reactions whose reaction equations are represented as
\begin{eqnarray}
A &\rightleftharpoons& X,\\
2X + Y &\rightleftharpoons& 3X,\\
X + B &\rightleftharpoons& Y + A.
\end{eqnarray}
Formally, the reaction set is defined as 
\begin{align}
        \R = \qty{
        \begin{aligned}
                r_1 &= (0,0,1,0)^\top \to (1,0,0,0)^\top,\\
    r_2 &= (2,1,0,0)^\top \to (3,0,0,0)^\top,\\
    r_3 &= (1,0,0,1)^\top \to (0,1,1,0)^\top,\\
    r_{-1} &= (1,0,0,0)^\top \to (0,0,1,0)^\top,\\
    r_{-2} &= (3,0,0,0)^\top \to (2,1,0,0)^\top,\\
    r_{-3} &= (0,1,1,0)^\top \to (1,0,0,1)^\top
        \end{aligned}
        }
\end{align}
where $\Z_{\geq 0}^\I$ is identified with $\Z_{\geq 0}^\abs{\I} = \Z_{\geq 0}^4$.
The kinetics $j$ is assumed to be mass-actional as follows.
\begin{align}
    j_1 &= k_1^+ [A],\\
    j_2 &= k_2^+ [X]^2[Y],\\
    j_1 &= k_3^+ [X][B],\\
    j_{-1} &= k_1^- [X],\\
    j_{-2} &= k_2^- [X]^3,\\
    j_{-3} &= k_3^- [Y][A],
\end{align}
where the notation $[Z]$ means the concentration of the chemical species $Z \in \I$.
The conventional flux and driving forces are defined as $J_m = j_m - j_{-m}$ and $F_m = \ln j_m - \ln j_{-m}$ for $m = 1,2,3$.
The system exhibits damped oscillations in concentration, kinetics and thermodynamic driving forces in \ref{fig:concentration}.

The S-graph $(\I, \E)$ of this system has an edge set 
\begin{align}
    \E = \qty{
    \begin{aligned}
        &X \gets Y, Y\gets X,\\
        &X \gets A, A\gets X,\\
        &Y \gets B, B\gets Y,\\
        &B \gets A, A\gets B.
    \end{aligned}
    },
\end{align}
whose reversibility is inherited from the original CRN.
The transition $\T$ for a conservation law $1$ has elements with
\begin{align}
\begin{split}
        \T_{XY} &= \qty( k_2^+ [X]^2 +  k_3^- [A]/2 ) [Y],\\ \T_{YX} &= \qty( k_2^- [X]^2 +  k_3^+ [B]/2 ) [X],\\
        \T_{XA} &= \qty( k_1^- + k_3^- [Y]/2 )[A],\\ \T_{AX} &= \qty( k_1^+ +  k_3^+ [B]/2 ) [X],\\
        \T_{XB} &= 0,\\ \T_{BX} &= 0,\\
        \T_{YA} &= 0,\\            \T_{AY} &= 0, \\
        \T_{YB} &= \qty( k_3^+ [X]/2 ) [B],\\            \T_{BY} &= \qty( k_3^- [A]/2 ) [Y] ,\\
        \T_{AB} &= \qty( k_3^+ [X]/2 ) [B],\\            \T_{BA} &= \qty( k_3^- [Y]/2 ) [A].
  \end{split}
\end{align}

\begin{figure}[htbp]
    \centering
    \includegraphics[width=.5\textwidth]{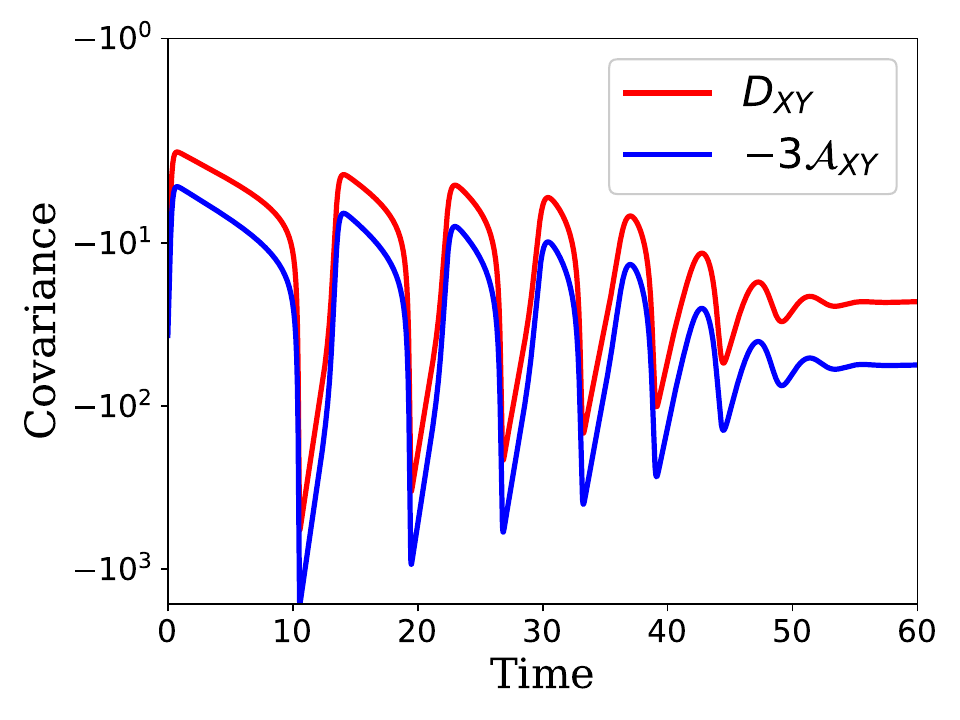}
    \caption{The time evolutions of the covariance $D_{XY}$ and the bound by the specieswise activities $\A_{XY}$.}
    \label{fig:covariance}
\end{figure}

The behaviour of the specieswise force $\F_{XY}$ and activity $\A_{XY}$ from $Y$ to $X$ and their bounds are observed.
As seen in the subsection \ref{specieswise forces}, $\F_{XY}$ could take values in the interval 
\begin{align}
    \Lambda_{XY} = \qty[\min_{r \in \qty{r_2,r_{-3}}} F_r, \max_{r \in \qty{r_2,r_{-3}}} F_r]
\end{align}
at each time. 
The time evolutions of $\F_{XY}$ and the possible interval $\Lambda_{XY}$ are showed in Fig \ref{fig:force_bound}.
In this simulation, we can comfirm that the area is actually $\Lambda_{XY} = [F_{r_{-2}}, F_{r_3}]$ for any time $t$.
The force and its upper bound is quite close at certain times, which is interpreted by the asymtotic approximation discussed in the last of the subsection \ref{specieswise forces}, together with the time series of the kinetics Fig \ref{fig:concentration} (b).
The time evolutions of the covariance $D_{XY}$ between $X, Y$ and the lower bound $-3\A_{XY}$ is seen in the Fig \ref{fig:covariance}.
Again, it can be seen that the two physical quantities are approximately coincident in situations where the largest reaction order, reaction 2, is dominant.

\begin{figure}[htbp]
    \centering
    \includegraphics[width=.5\textwidth]{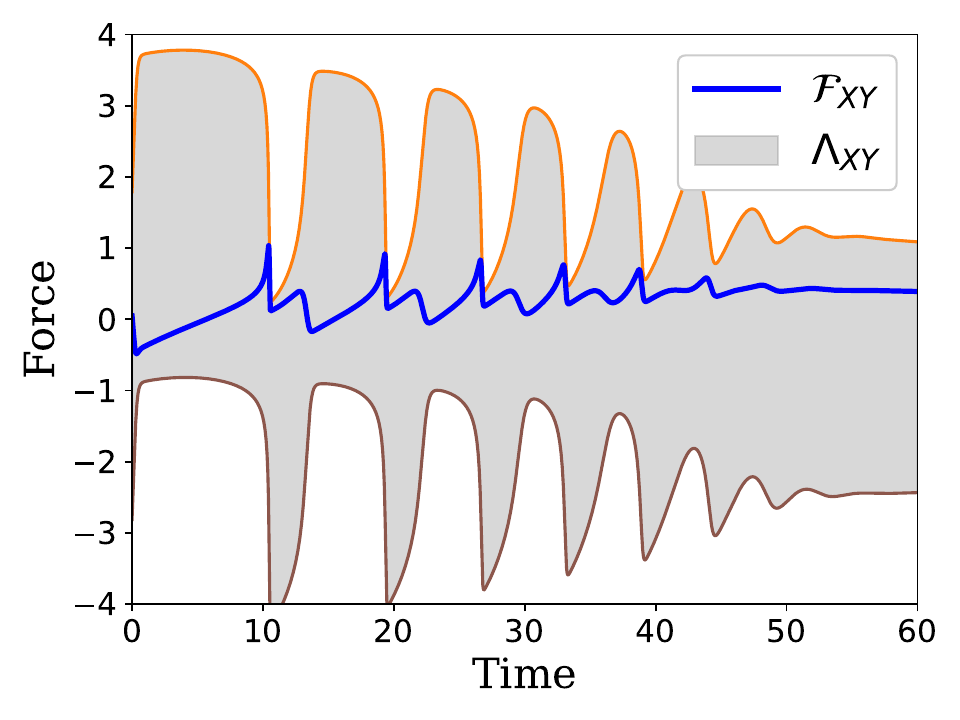}
    \caption{The time evolutions of the specieswise force $\F_{XY}$ from $Y$ to $X$ and its possible interval $\Lambda_{XY}$. The upper and lower line in the figure represent $F_{r_2}$ and $F_{r_{-3}}$, respectively.}
    \label{fig:force_bound}
\end{figure}

By observing the numerical results, two properties can be identified for our defined quantities below.
First, we argue that the physical quantities we have defined are properly interpretable.
Indeed, we have so far confirmed both theoretically and numerically that specieswise forces and activities are related to conventional driving forces and covariance.
In particular, for the former, Figure \ref{fig:concentration} (a) shows that $X$ increases and $Y$ decreases in the long term, which is consistent with a positive $Y$ to $X$ force for most of the time $0\leq t\leq 60$. 
Secondly, the specieswise force $\F_{XY}$ and activity $\A_{XY}$ is found to inherit the oscillatory nature of the system in the simulation.

\section{conclusion and discussion}
This paper introduced a novel framework for analyzing CRN dynamics by defining "transitions" as a new physical quantity and reinterpreting the dynamics on a species graph. The dynamics on this graph were proven conservative by scaling concentrations according to the network's conservation laws. Thermodynamic quantities like forces and activities were then formulated at the species level based on these transitions and the relation between the specieswise thermodynamic quantities and conventional thermodynamic quantities were investigated. When applied numerically to the Brusselator oscillating reaction system, the specieswise quantities exhibited intuitive behavior aligned with theoretical predictions. This transition-based graphical approach provides an alternative perspective for understanding the thermodynamics of complex CRNs with potential for enhanced modeling and analysis. Further research is merited into the broader implications and applications of this specieswise thermodynamic formulation.

One potential limitation of our framework is that the newly introduced concept of "transitions" may not precisely mirror the actual physical transformations occurring at the molecular level. To illustrate, consider a second-order chemical reaction $A+B \to A' + B'$ through which the physical chemistry explains the molecules $A$ and $B$ undergo structural changes to become $A'$ and $B'$ respectively. Following this depiction, the reaction should solely contribute to the transitions $A\to A'$ and $B\to B'$. However, our methodology additionally accounts for the transitions $A\to B'$ and $B\to A'$, which do not directly correspond to the molecular rearrangements. Despite this potential disconnect from microscopic mechanisms, our transition-based approach offers the significant advantage of being systematically computable from the system's stoichiometric and kinetic information, lending itself to practical analysis workflows.

Looking ahead, we anticipate our framework enabling deeper dynamical analysis of reaction networks, particularly metabolic systemsand others central to cellular processes. Theoretically, exploring how stochastic thermodynamics laws manifest in CRNs under our specieswise formulation is intriguing. For instance, it is an interesting topic that how the relations between correlations and driving forces at a steady state for Markov jump system \cite{PhysRevLett.131.077101} potentially transform in the CRN context. As this specieswise approach develops, prospects include analyzing complex network dynamics and bridging disparate fields through a unified reaction network lens, with potential for significant discoveries.


\newpage
\bibliography{article}

\end{document}